\documentclass[twocolumn,prb]{revtex4}
\usepackage{natbib}
\usepackage{graphicx}
\usepackage{subfigure} 
\usepackage[latin1]{inputenc}
\usepackage[brazil]{babel}
\usepackage{amsmath}
\usepackage{setspace}
\usepackage{amssymb}
\usepackage{color}

\begin{document}

\title{The Quantum Spherical Spin Glass Model: A Limitation to Static Approximation.}
\author{Vilarbo da Silva Junior}
\email{vilarbos@unisinos.br}
\author{Alexsandro M. Carvalho}
\email{alexsandromc@unisinos.br}
\affiliation{Centro de Ciências Exatas e Tecnológicas, Universidade do Vale do Rio dos Sinos, Caixa Postal 275, 93022-000 São Leopoldo RS, Brazil}

\begin{abstract}
In this work we confront the static approximation with a exact solution in the quantum spherical p-spin interaction model ($p\rightarrow \infty$ and $p=2$). This study indicates that the static approximation corresponds to exact solution in the cases $p\rightarrow \infty$ and $p = 2$ in the classic regime. On the other hand, it differs from the exact solution for $p=2$ in the quantum regime.
\end{abstract}
\maketitle

\section{Introduction}
The term Spin Glass (SG) appeared to designate a class of metallic alloys which are formed from noble metals ions weakly diluted in magnetic transition metals. One reason for the name spin glass is because the magnetic moments of these alloys present a locally fixed orientation without any periodic ordering, so conceptually it is similar to amorphous structures as conventional glass. In other words, the spin glass phase can be understood as a set of spins exhibiting a frozen phase at low temperatures without magnetic long range order.

The first theoretical model developed to study the existence of a spin glass phase was proposed Edwards and Anderson~\cite{Edwards1975}. Subsequently, a version of the infinite-range Edwards-Anderson model was developed by Sherrington and Kirkpatrick~\cite{Kirk1978} (SK model). Unlike pair interactions, proposed in the two previous models, Crisanti and Sommers~\cite{Crisanti1992} developed a model that generalizes the interactions for p spins, know as classical spherical (CS) p-spin interaction spin glass model.

However, at low temperature, experimental evidence suggests that quantum effects are significant~\cite{Kenning1991}. Moreover, quantum theoretical models were developed to describe the SG phase at this temperature regime. In particular, we cite the quantum spherical (QS) spin glass model~\cite{Shukla1981,Ye1993}.

Based on previous model, Cugliandolo et al.~\cite{Cugliandolo2001} conducted a formal study of QS p-spin interaction model. Unfortunately, it was not possible to treat this model analytically for a generic value of p. Thus, after performing the replica method, the authors used a static approximation (SA) ~\cite{Bray1980}. However, Menezes and Theumann~\cite{Menezes2007} showed that the effective action of the quantum spherical model for $p = 2$ is invariant over a generalized form of  Becchi-Rouet-Stora-Tyutin supersymmetry, and thus a result via annealed average is exact for this model.

Within this context, in this paper, we compare the static approximation with the exact solution for the QS p-spin interaction model in two particular cases: (i) $p=2$ and (ii) $p\rightarrow \infty$. The objective of this study is to quantify how good is the static approximation for the values of p indicated. For this purpose, we following the standard procedure of replica method~\cite{Kirk1978} considering the hypothesis of replica symmetry (RS), which is sufficient for this model~\cite{Theumann2011}. Our results suggest that the limit $p \rightarrow \infty$ the static approximation is exact. On the other hand, the case $p=2$ the static approximation is exact in the classic limit but is not good in the quantum limit.

This paper is organized as follows: In Sec.II and IIIA, respectively, we review the QS p-spin interaction model and exact solution for $p=2$. In Sec. IIIB, we show the exact solution for $p\rightarrow \infty$. In Sec. IV, we solve the QS model using static approximation for $p=2$ and $p\rightarrow \infty$. In Sec. V, we compare the exact solution with the result obtained via static approximation. Finally, in Sec. VI, we discuss the limitation of the result obtained via static approximation for QS p-spin interaction model.

\section{Model}
The hamiltonian for QS p-spin interaction model is given by
\begin{eqnarray}
\hat{\mathcal{H}}=&&\frac{1}{2I}\sum_{i=1}^{N}\hat{P}_{i}^{2}-\sum_{1\leq i_1 < \cdots  < i_p \leq N} J_{i_1\cdots i_p}\hat{S}_{i_1}\cdots\hat{S}_{i_p} \nonumber \\ &+&\mu\sum_{i=1}^{N}\hat{S}_{i}^{2},
\label{eq1}
\end{eqnarray}
where the spin operators have continuous eigenvalues $S_{i} \in (-\infty,\infty)$, $\hat{P}_{i}$ is the momentum operator canonically conjugated to $\hat{S}_{i}$ (it satisfies the relation $[\hat{S}_{k},\hat{P}_{l}]=i\delta_{kl}$), $I$ is the moment of inertia of the spins (quantum rotors), $\mu$ is the Lagrange's multiplier for the mean spherical constraint ($\sum_{i=1}^{N}\left<\hat{S}_{i}^2\right>=N$) and $J_{i_1\cdots i_p}$ are the elements of a random symmetric matrix (distributed according to a Gaussian with zero mean and variance $\sigma^2 = J^2p!/(2N^{p-1}$)). Since the distribution is the same for any set of $p$ spins, this is equivalent to the infinite range (mean field) - as in the SK model.

\section{Exact Solution}
\label{secExactSol}
To solve the system means, in the context of this work, to find an analytical form for the grand thermodynamical potential, in the limit of replica's number $n\rightarrow 0$ and for a macroscopical system ($N\rightarrow \infty$), from which can be obtained physical properties of the system. For this end, we solve the integrals that appear in the replicate partition function and obtaining explicit forms to determine the fields (in the context of field theory) from the saddle point equations.

To obtain the partition function of the model we follow the Feynman's prescription to path integrals~\cite{Feynman1965} with the hamiltonian in~(\ref{eq1}). The connection with thermodynamics is made through the grand thermodynamic potential $\Omega$ ($\propto \log{Z(\textbf{J})}$). We can make an analogy between the parameter $\mu$ of spherical constraint and the chemical potential. The randomness is treated by realizing the configurational average in $\log{Z(\textbf{J})}$ (quenched) and, in order to accomplish it, we use the replica method. Finally, to uncouple the imaginary times (from the Feynman's path integral)  we introduce the Fourier series for quantities dependent on these times. After realize all these steps, the grand thermodynamic potential becomes
\begin{equation}
\frac{\beta \Omega}{N}=\lim_{n\rightarrow 0}\frac{1}{n}\,G[\textbf{q}(\omega_{m})],
\label{eq2}
\end{equation}
where
\begin{eqnarray}
G[\textbf{q}(\omega_{m})]&=&-\frac{J^{2}\beta}{4}\sum_{\alpha\,\nu}\int_{0}^{\beta}d\tau\left[\sum_{m}e^{-i\,\omega_{m}\tau}q_{\alpha\,\nu}(\omega_{m})\right]^{p}\nonumber\\
&-&\frac{1}{2}\left(\sum_{m}\log{[D_m]}+\sum_{\alpha,m}\lambda_m q_{\alpha\,\alpha}(\omega_{m})\right),
\label{eq3}
\end{eqnarray}
$D_m = \det{(\textbf{q}(\omega_{m}))}$ and $\lambda_m = I\omega_{m}^{2}\beta+2\beta\mu$. The term $\textbf{q}(\omega_{m})$ is the overlap's matrix between replicas, $\omega_{m}=2\pi\,m/\beta$ are the Matsubara frequencies~\cite{Negele1988} for bosons (since the variables has one commutation algebra). The fields $q_{\alpha\,\nu}(\omega_{m})$ are determined by $\delta G[\textbf{q}(\omega_{m})]/\delta q_{\alpha\,\nu}(\omega_{m})=0$ in the Eq.~(\ref{eq3}), results
\begin{equation}
\frac{\beta J^{2}p}{2}\int_{0}^{\beta}d\tau\, e^{i\,\omega_{m}\tau}q_{\alpha\,\nu}^{p-1}(\tau)+[\textbf{q}^{-1}(\omega_{m})]_{\alpha\,\nu}=\lambda_m\delta_{\alpha\,\nu}.
\label{eq4}
\end{equation}

In the next subsection, we study the Eq.~(\ref{eq4}) for two specific cases: $p=2$ and $p\rightarrow \infty$.

\subsection{Case $p=2$}
\label{subsecExactSolp2}

In the case $p=2$ the replica symmetric \textit{ansatz}, parametrizing $q_{\alpha\,\nu}(\omega_{m})=(q_{o}(\omega_{m})-q(\omega_{m}))\delta_{\alpha\,\nu}+q(\omega_{m})$, is sufficient for a complete description of the model. This hypophyses allows us obtain explicitly the elements of the inverse matrix $\textbf{q}^{-1}(\omega_{m})$.

The $q(\omega_{m})=0$ solution is always valid (paramagnetic solution) for the saddle-point equations. With this solution we obtain a quadratic equation for $q_{o}(\omega_{m})$, which solution is
\begin{equation}
q_{o}(\omega_{m})=\frac{\lambda_m-\sqrt{\lambda_m^2-4(\beta J)^{2}}}{2(\beta J)^{2}}.
\label{eq5}
\end{equation}

The spherical constraint can be written as $\sum_{m}q_{o}(\omega_{m})=1$, and the sum over frequencies can be solved using a standard procedure with integrals in the complex plane~\cite{Negele1988}. This approach leads us to
\begin{equation}
\frac{1}{2\pi J^{2}}\int_{\frac{L_{-}}{\sqrt{I}}}^{\frac{L_{+}}{\sqrt{I}}}dx H(x)\coth{\left(\frac{\beta x}{2}\right)}=1,
\label{eq6}
\end{equation}
where $H(x) = \sqrt{4J^{2}-(2\mu-Ix^{2})^{2}}$ and $L_{\pm}=\sqrt{2(\mu\pm J)}$. This integral over the real variable $x$ is well defined if $\mu \geq J$. Thus, $\mu$ ``sticks'' at the value $\mu_{c}=J$ below a certain temperature $T_{c}$ (critical temperature) obtained in Eq.~(\ref{eq6}) by sitting $\mu=J$ for each fixed $I$. The critical value $I_{c}^{Exact}$ (where $T_{c}=0$) is obtained analytically as $1/JI_{c}^{Exact}=\frac{9\pi^{2}}{16}\simeq 5.5$.

\subsection{Case $p\rightarrow \infty$}
\label{subsecExactSolpInfty}

Now we treat the model in the limit $p\rightarrow \infty$. For a classical Ising spin case this limits corresponds to random energy model~\cite{Derrida1981,Gross1984} and a quantum model with Ising spins in the presence of a transverse field is treated by Obuchi, Nishimori and Sherington~\cite{Obuchi2007}.

Assuming replica symmetry in Eq.(\ref{eq4}) we see that $q(\omega_{m})=0$ is always a solution. But, when $p\rightarrow \infty$ it follows that $p\,q^{p-1}(\tau)\rightarrow 0$ (if $0\leq q(\tau)<1$) so $q(\omega_{m})=0$. Now, for $p\,q^{p-1}(\tau)\rightarrow 1$ (if $ q(\tau)=1$) and then $q(\omega_{m})\rightarrow \infty$ (unphysical solution) or $q_{o}(\omega_{m})=q(\omega_{m})=1$ which can not occur at a finite temperature.
Therefore, within the limit $p\rightarrow \infty$ the system presents only the paramagnetic phase. Setting $q(\omega_{m})=0$ in the saddle point Eq.~(\ref{eq4}) for $q_{o}(\omega_{m})$ (with $\alpha=\nu$), we get
\begin{equation}
\frac{\beta J^{2}p}{2}\int_{0}^{\beta}d\tau\, e^{i\,\omega_{m}\tau}q_{o}^{p-1}(\tau)+\frac{1}{q_{o}(\omega_{m})}=\lambda_m.
\label{eq7}
\end{equation}
Thus for $p\rightarrow \infty$, we have $pq_{o}^{p-1}(\tau)\rightarrow 0$ (if $q_{o}(\tau)<1$), thus $q_{o}(\omega_{m})=(I\omega_{m}^{2}\beta+2\beta\mu)^{-1}$.

Using the procedure made in Sec.~\ref{subsecExactSolp2}), we find
\begin{equation}
\frac{1}{2}\sqrt{\frac{1}{2\,\mu\, I}}\coth{\left[\beta \sqrt{\frac{\mu}{2I}}\right]}=1.
\label{eq8}
\end{equation}

This equation allows us get $\mu(T)$ as a function of the temperature for a fixed $I$. So, $\mu=0$ is not allowed, i.e., $T_{c}=0$ for all $I$.

\section{Static Approximation}
\label{secSatAproxSol}

To employ the formalism of Feynman's path integrals, we obtain a functional integral dependent on the imaginary time ($\tau$) and its associated fields have the same dependence. This adds a great degree of difficulty in the analytical viewpoint and it is not possible to advance in the problem for any $p$ (see Eq.~(\ref{eq4}) for example).

In order to circumvent this issue, Bray and Moore~\cite{Bray1980} proposed an approximate method, which is referred as the static approximation (SA) and consists in neglect the (imaginary) time dependence of the order parameters (fields).

Following the proposal of Obuchi, Nishimori and Sherrington~\cite{Obuchi2007} to implement the static approximation
(combined with RS \textit{ansatz}), we get the grand thermodynamic potential $\Omega_{SA}$ as
\begin{eqnarray}
\frac{\beta \Omega_{SA}}{N}&=&-\frac{(\beta J)^{2}}{4}(q_{o}^{p}-q^{p})-\frac{1}{2}\log{(q_{o}-q)}\nonumber\\
&-&\frac{q}{2(q_{o}-q)}+\log{\left[2\sinh{\left[\frac{\beta}{2}\sqrt{\frac{2\mu}{I}}\right]}\right]}\nonumber\\
&-&\frac{1}{2}\log{(2\beta\mu)}+\beta\mu q_{o}.
\label{eq9}
\end{eqnarray}
It is analogous to eq.~(\ref{eq2}) and~(\ref{eq3}), where $\Omega_{SA}$ denotes the grand canonical potential over SA. Now, the saddle point equations $\partial \Omega_{SA}/\partial q_{o}=\partial \Omega_{SA}/\partial q=0$, together with the mean spherical condition $\partial \Omega_{SA}/\partial \mu=N$ leads us to relations similar to eq.~(\ref{eq4}), more precisely:
\begin{equation}
 \left \{  \begin{array}{lll}  \frac{( \beta J)^{2}}{2}pq^{p-1}-
     \frac{q}{(q_{0}-q)^{2}} &=&0; \\
 \frac{( \beta J)^{2}}{2}pq_{0}^{p-1}+ \frac{q_{0}-2q}{(q_{0}-q)^{2}}
 &=&2 \beta \mu; \\
1+ \frac{1}{2 \beta \mu}- \frac{1}{2} \sqrt{ \frac{1}{2 \mu I}} \coth{
  \left[ \beta \sqrt{ \frac{ \mu}{2I}} \right]}&=&q_{0}.
 \end{array}  \right.
 \label{eq10}
 \end{equation}

In the next subsection we explore solutions for the system~(\ref{eq10}) in the same values of $p$ treated in Sec. III.
\subsection{Case $p=2$}
\label{subsecSASolp2}

 Replacing $p=2$ in the saddle point equation~(\ref{eq10}) is relatively simple. We note that the paramagnetic solution ($q=0$) is always possible and it is sufficient to describe the phase diagram. So, making $q=0$ in the second equation in (\ref{eq10}), we come to a quadratic equation in the variable $q_{o}$, which solution is given by
\begin{equation}
q_{o}=\frac{\left(\frac{\mu}{J}\right)-\sqrt{\left(\frac{\mu}{J}\right)^{2}-1}}{(\beta J)}.
\label{eq11}
\end{equation}
The equation above is analogous to eq.~(\ref{eq5}) for $\omega_{m}=0$. Furthermore, in order that $q_{o}$ is real, the condition $\mu \geq J$ must be satisfied. Thus, we conclude that $\mu_{c}^{SA}=J$ for $T\leq T_{c}^{SA}$, where $T_{c}^{SA}$ is the critical temperature below which $\mu$ sticks at $J$ over the SA. This is the same value from sticks for $\mu$ in the Sec.~\ref{subsecExactSolp2}, when we treat the model exactly.

In the critical temperature $\mu=\mu_{c}^{SA}=J$, and then, setting it in eq.~(\ref{eq11}) we get $q_{o}=q_{o}^{c}=T_{c}^{SA}/J$. So, with this information in the spherical condition (third equation in (\ref{eq10})) we obtain the equation that gives us $T_{c}^{SA}$ for a fixed $I$ and, consequently, the phase diagram, which is
\begin{equation}
\frac{1}{2}\left(\frac{T_{c}^{SA}}{J}\right)=1-\frac{1}{2}\sqrt{\frac{1}{2J\,I}}\coth{\left[\left(\frac{J}{T_{c}^{SA}}\right)\sqrt{\frac{1}{2J\,I}}\right]}.
\label{eq12}
\end{equation}

Finally, the critical value of $I$ ($I_{c}^{SA}$) can be obtained analytically by taking the limit $T_{c}^{SA}\rightarrow 0$ in eq.~(\ref{eq12}), where we get $1/JI_{c}^{SA}=8$.

\subsection{Case $p\rightarrow \infty$}
\label{subsecSASolpInfty}

In this section, we approach the QS $p$-spin interaction model in the limit $p\rightarrow \infty$ in the Static Approximation. In this case the model becomes extremely simple in the analytical viewpoint and it is found only in paramagnetic phase as in Sec.~\ref{subsecExactSolpInfty}. Indeed, the first equation in (\ref{eq10}) always admits as solution $q=0$, characterizing the paramagnetic phase. But, if we take the limit $p\rightarrow \infty$ it follows that $p q^{p-1}\rightarrow 0$ (if $0\leq q<1$) whence $q=0$. Additionally, $p q^{p-1}\rightarrow \infty$ (if $q=1$), thence $q\rightarrow\infty$ (unphysical solution) or $q_{o}=q=1$ that may not occur for a finite temperature. Therefore we conclude that $q = 0$ is the only admissible solution. Now we return to the second equation in (\ref{eq10}) with $q=0$ and we obtain an expression similar to eq.~(\ref{eq7}):
\begin{equation}
\frac{(\beta J)^{2}}{2}p\,q_{o}^{p}-2\beta\mu q_{o}+1=0.
\label{eq13}
\end{equation}
Now, taking the limit $p\rightarrow \infty$, we have $pq_{o}^{p}\rightarrow 0$ (if $q_{o}<1$) this leads us to $q_{o}=\frac{1}{2\beta\mu}$. Therefore, with $q_{o}=\frac{1}{2\beta\mu}$ in the third Eq.~(\ref{eq10}), we get
\begin{equation}
\frac{1}{2}\sqrt{\frac{1}{2\,\mu\, I}}\coth{\left[\beta \sqrt{\frac{\mu}{2\,I}}\right]}=1,
\label{eq14}
\end{equation}
which is exactly the same obtained in the exact treatment (eq.~(\ref{eq8}) .

\section{Static Approximation versus Exact Solution}
\begin{figure}[!h]
  \includegraphics[width=8cm]{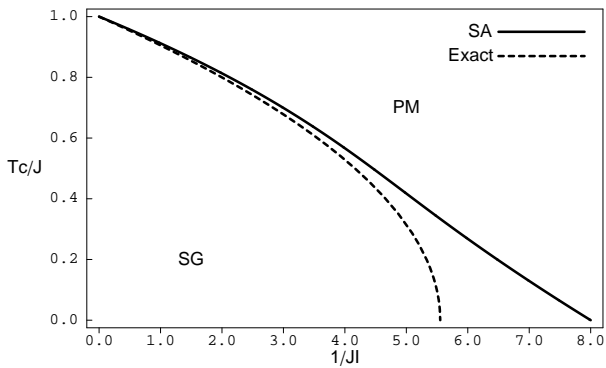}\\
  \caption{Phase diagram for $p=2$. The dotted (Exact) and continuous (SA) lines result from the numerical solution of Eq.~(\ref{eq6}) and~(\ref{eq12}), respectively. The paramagnetic phase (PM) exists above the coexistence curve, whereas the spin glass phase (SG) below it.}\label{fig:PD}
\end{figure}

In the two previous sections we indicate the exact and approximate (SA) solutions for the QS p-spin interaction model.To compare the solutions obtained for p = 2, we find the numerical solutions for the eqs.~(\ref{eq6}) and~(\ref{eq12}) (see Fig.~\ref{fig:PD}).
According to this phase diagram, we can see that the static approximation is good in the classical limit (small $1/JI$), however it does not adequately describe our model when the quantum effects become relevant. Additionally, we can check that the critical values $I_c$, both exact and approximate, agree well with the estimated values in Fig.~\ref{fig:PD} ($1/JI_{c}^{Exact} \simeq 5.5 $ and $1/JI_{c}^{SA}=8$). The curve $T_{c}/J$ versus $1/JI$ separating the spin glass (SG) and paramagnetic (PM) phases.

For $p\rightarrow \infty$, the comparison between these two solutions is immediate (Eq.~(\ref{eq8}) and~(\ref{eq14})) and $T_c = 0$ for both. In others words there is no phase transition and only the paramagnetic phase is present. Thus, the solution given by SA corresponds to the exact solution.
\vspace{3cm}
\section{Conclusions}
In this work we compare the static approximation with exact solution for the quantum spherical p-spin interaction model for two particular cases: $p=2$ and $p\rightarrow \infty$.

For $p=2$ our results indicate that the correspondence between the static approximation and the exact solution depends directly on the value of the moment of inertia $I$. In other words, when $I \rightarrow \infty$ (classic limit) the static approximation is exact. However, for $I \rightarrow 0$ the exact solution differs considerably from the static approximation. This discrepancy is linked to the fact that to obtain the exact critical temperature of the system was necessary to accomplish a sum over all frequencies, while in the static approximation is considered single frequency $\omega_m = 0$.  This limitation to static approximation in the quantum regime was also observed in the SK model with a transverse field~\cite{thirumalai1989}.

For $p\rightarrow \infty$ we find that the static approximation is exact, since there is no phase transition. This result is in agreement with the obtained to p-spin interaction model with ferromagnetic bias and transverse field~\cite{Obuchi2007}.

Finally, our results suggest that static approximation is a valid starting point to treat quantum spin glass models, however, it needs improvements to completely describe this class of models.

\begin{acknowledgments}
We thank Professor Alba Theumann (\textit{in memoriam}) for valuable lessons and Professor Rogerio Steffenon for the frequent incentive. We also thank José Luiz Ferreira Jr for a critical reading of the manuscript and useful suggestions.
\end{acknowledgments}

\bibliographystyle{phcpc} 

\end{document}